\documentclass[fleqn,twoside]{article}
\usepackage{gc}
\usepackage{latexsym,amsmath}
\renewcommand{\theequation}{\thesection\arabic{equation}}
\newcommand{\be}{\begin{equation}}
\newcommand{\ee}{\end{equation}}
\newcommand{\ben}{\begin{enumerate}}
\newcommand{\een}{\end{enumerate}}
\newcommand{\beqs}{\begin{eqnarray}}
\newcommand{\eeqs}{\end{eqnarray}}

\newcommand{\la}{\label}

\newcommand{\bmin}{\begin{minipage}}
\newcommand{\emin}{\end{minipage}}

\newcommand{\ag}{\alpha}
\newcommand{\La}{\Lambda}
\newcommand{\lam}{\lambda}

\newcommand{\m}{\mu}
\newcommand{\n}{\nu}

\newcommand{\vf}{\varphi}

\newcommand{\pdr}{\partial}

\newcommand{\lp}{\left(}
\newcommand{\rp}{\right)}
\newcommand{\lb}{\left[}
\newcommand{\rb}{\right]}

\newenvironment{definicion}[1]{\begin{flushleft} $\bullet$ {\bf \Large #1 \hfill}}{\end{flushleft}}
\newcommand{\bedef}{\begin{definicion}}
\newcommand{\eedef}{\end{definicion}}

\newcommand{\rmd}{\mbox{d}}
\newcommand{\rme}{\mbox{e}}

\heads{J. Quiroga \& Y. Shaido}
      {Quantum stabilization of dilatonic Anti-de Sitter Universe}

\begin{document}
\twocolumn[\Arthead{8}{2002}{4 (32)}{1}{4}

\Title{Quantum stabilization of dilatonic Anti-de Sitter Universe}

\Author{John Quiroga Hurtado\foom 1 \foom 2}
       {Department of Physics , Universidad Tecnologica de
Pereira, Colombia}{and Lab. for Fundamental Study, Tomsk State
Pedagogical University, Tomsk
634041, Russia}\\

\Author{Yulia A. Shaido\foom 3 }
       {Lab. for Fundamental Study, Tomsk State Pedagogical University, Tomsk
634041, Russia}

\Abstract{The possibility of quantum creation of a dilatonic AdS
Universe is discussed.Without dilaton it is known that quantum
effects lead to the annihilation of AdS Universe. We consider the
role of the form for the dilatonic potential in the quantum
creation of a dilatonic AdS Universe. Using the conformal anomaly
for dilaton coupled scalar, the anomaly induced action and the
equations of motion are obtained. The anomaly induced action is
added to classical dilaton gravity action. The solutions of the
full theory which correspond to quantum-corrected AdS Universe are
given for number of dilatonic potentials.}
          ]

\email 1 {jquiroga@tspu.edu.ru} \email 2{jquiroga@utp.edu.co}
\email 3{Shaido@mail2000.ru}

\section{Introduction}
There is much interest now to various studies of Anti-de Sitter
(AdS) space. This is motivated by AdS/CFT correspondence (for a
review, see \cite{review}) according to which (in its simplest
form), the properties of classical AdS space may correspond to the
ones of some dual conformal field theory in less dimensions.

From another side, quantum field theory in curved spacetime (see
\cite{BOS}, for a review) may help in our understanding of the
origin of the early Universe. In particular, one of most
successful inflationary models (so called trace anomaly induced
inflation \cite{Starobinskyetc}) is based on quantum creation of
de Sitter Universe by matter quantum effects. It is very
interesting to understand if this is specific property of de
Sitter space or it is quite general phenomenon?

In this work, we consider the role of dilaton coupled quantum
matter to stabilization of AdS Universe. It is known that without
dilaton, quantum effects destabilize and annihilate AdS Universe
\cite{brevik}. These quantum effects in our work are accounted for
by four-dimensional dilaton conformal anomaly (for a review, see
\cite{odintsov}). It is very interesting to note that such
dialaton conformal anomaly in $N=4$ super Yang-Mills theory with
$N=4$ conformal supergravity may have the holographic origin (via
AdS/CFT) as it was shown in \cite{SN}.

Using anomaly induced effective action for dilaton coupled scalar
and adding it to the classical gravity action, we analyze the
equations of motion, for a number of dilatonic potentials. It is
shown that the possibility of quantum creation of dilatonic AdS
space occurs.

\section{Description of the Model}

First of all we will consider quantum fields in AdS space. At the
end of this section we will also review quantum annihilation of
Anti-de Sitter Universe \cite{brevik}.

Our model is given as a 4-dimensional Anti-de Sitter spacetime
$(AdS_4)$ and the metric is chosen to be \cite{brevik}
\beqs
 ds^2=e^{-2\lambda
\tilde{x_3}}(dt^2-(dx^1)^2-(dx^2)^2)-(d\tilde{x}^3)^2\,, \la{metr}
\eeqs
with a negative effective cosmological constant $\La = -\lambda
^2$. We may rewrite this metric in conformally flat form  by
using the following transformation

\be y= x^3=\frac{e^{-\lambda \tilde{x_3}}}{\lambda}. \la{eq2}\ee

taking $a=e^{-\lambda \tilde{x_3}}=1/(\lam x^3)=1/(\lam y)$, we
finally obtain for the metric the following expression
\begin{equation}
ds^2=a^2(dt^2-dx^2)= a^2\eta_{\m\n}dx^\mu dx^\n .
\label{metric}\end{equation}

It is important to remark that the obtained form for the metric is
the most useful in the study of quantum gauge theory via the SG
dual.

Now, let us imagine that the early Universe is filled by some
grand unified theory (GUT). In that case, it is known \cite{BOS}
that it is enough to consider only free fields, as radiative
corrections are not essential. It is also important to remark that
unlike to the de Sitter space the AdS space is supersymmetric
background for GUT in the case if it is SUSY. For a review of the
quantum fields on negative curvature space see ref. \cite{sergio}.

In order to construct the effective action in terms of the quantum
scalar field theory, (for simplicity) we will consider the
anomaly, expressed as (for a review see \cite{Duff94,odintsov})
\beqs\!\!\!\!\!\!\!\!\!\!\!\!
T&=&b\left( F+{2\over 3}\,\Box R\right)+b'G+b''\,\Box R\,+\nonumber\\
\!\!\!\!\!&&+a_1\frac{[(\nabla f)(\nabla f)]^2}{f^4}+a_2 \Box
\left(\frac{(\nabla f)(\nabla f)}{f^2}\right), \la{anom}\eeqs

where for single scalar,
\beqs b&=&\frac{1}{120(4\pi)^2}\;,\qquad
b'=-\frac{1}{120(4\pi)^2}\;,\nonumber\\
a_1&=&\frac{1}{32(4\pi)^2}\;\;,\qquad a_2=\frac{1}{24(4\pi)^2}\;,
\eeqs

while $F$ is the square of the Weyl tensor in four dimensions ,
and $G$ is the  Gauss-Bonnet invariant
\be
F=R_{\mu\nu\alpha\beta}R^{\mu\nu\alpha\beta}-2R_{\mu\nu}R^{\mu\nu}
+{1\over 3}\,R^2\;, \ee

Note that the  constant $b''$ is an arbitrary parameter (it can be
changed by a finite renormalization of the gravitational action).
As the physical meaning does not change we may put  $b''=0$. The
terms with coefficients $a_1$ and $a_2$ correspond to the
contribution of the dilaton, where $f$ is an arbitrary dilatonic
coupling  $f=f(\varphi)$. Their contribution to four dimensional
conformal anomaly is found in ref.\cite{SNO}. The initial action
for dilaton coupled conformal quantum scalar is given by:
\beqs S=\int\rmd^4\!x\sqrt{-g}\phi f(\Box-\frac{1}{6}R)\phi\;,
\eeqs
where $\phi$ is quantum scalar filed.

Note also that if in (\ref{anom}) the terms, which contain
dilaton, are omitted  we obtain the well known conformal anomaly,
for scalar field.

\section{Effective action and equations of motion}
By using this conformal anomaly it is not difficult to construct
the anomaly-induced effective action \cite{ReigertTseytlinetc}.
Taking into account that it is useful to consider a conformal
metric for a AdS space, one may write the metric for our case as
$g_{\mu\nu}=\rme^{2\sigma(y)}\eta_{\mu\nu}$, which is similar to
(\ref{metric}), where $\eta_{\mu\nu}$ is the Minkowski metric.
After this, using the techniques of ref.
\cite{ReigertTseytlinetc}, the following anomaly-induced effective
action on AdS space is obtained,
\beqs
\!\!W\!\!\!\!\!&=&\!\!\!\!\!\int\rmd^4\!x{\Big\{2b'\sigma\Box^2\sigma
-\frac{1}{12}\lp b''+\frac{2}{3}(b+b')\rp\Big(6\Box\sigma}+\nonumber\\
& &+6\eta^{\mu\nu}
(\partial_{\mu}\sigma)(\partial_{\nu}\sigma)\Big)^2+ a_1\frac{[(\nabla f)(\nabla f)]^2}{f^4}\sigma+\nonumber\\
\!\!\!\!\!\!& &+a_2 \Box \left(\frac{(\nabla f)(\nabla
f)}{f^2}\right)\sigma + a_2\frac{(\nabla f)(\nabla
f)}{f^2}[(\nabla\sigma)(\nabla\sigma)] \Big\}. \nonumber\eeqs
 Since $\sigma$ depends only on  $y$ and supposing that
$f=\varphi$, this expression may be simplified, and finally the
following expression is obtained for the effective action:
\beqs\!\!\!\!\!\!
 W\!\!\!\!\!\!&=&\!\!\!V_3\!\!\!\int\!\!\!\rmd
y{\Big[2b'\sigma\sigma''''
-3\big(b''+\frac{2}{3}(b+b')\big)(\sigma''+(\sigma')^2)^2 +}\nonumber\\
\!\!\!\!\! &+&\!\!\! a_1\frac{(\varphi')^4}{\varphi^4}\sigma +
a_2\lb\frac{(\vf')^2}{\vf^2}\rb''\!\!\!\sigma +
a_2\frac{(\vf')^2}{\vf^2}(\sigma')^2\Big]. \eeqs
In the last equation we have \, $\sigma'=\rmd\sigma/\rmd y$. On
the other hand it is well known that the total effective action
consists of $W$ plus some conformally invariant functional. But,
since we are considering a conformally flat background, this
conformally invariant is a non-essential constant. It could become
more important, similar to a kind of Casimir energy, if we
considered periodicity on some of the coordinates (say, AdS BH)
because in such a situation it would depend on the radius of the
compact dimension.

The quantum matter effects in the AdS Universe may be accounted by
adding the anomaly-induced action to the classical gravitational
action, which in this case includes a dilatonic potential $V(\vf)$
and a kinetic term for the dilaton, and  has the following form,
\beqs\la{scl}
S_{\rm cl}&=&-{1\over \kappa}\int\rmd^4\!x\,\sqrt{-g}\,(R+\frac{\beta}{2}g^{\m\n}\pdr_\m\vf\pdr_\n\vf+V(\vf)+\nonumber\\
& & +6\Lambda)=\nonumber\\
&=&-{1\over \kappa}\int\rmd^4\!x\,\rme^{4\sigma}
(6\rme^{-2\sigma}((\sigma')^2+(\sigma''))+\nonumber\\
& & + \frac{\beta}{2}\rme^{2\sigma}(\vf')^2+V(\vf)+6\Lambda)\;.
\eeqs
Now, in order to describe the dynamics of the whole quantum
system, we must take the sum of the classical action and the
quantum effective action.

Varying the action $S_{cl}+W$ with respect to $\sigma$ and $\vf$,
one may obtain the equations of motion, assuming in this case that
$\sigma$ and $\vf$ depends only on conformal coordinate $y$, we
finally arrive at the following equations of motion
\beqs\la{ecmov}\!\!\!\!\!\!\!\!\!\!\!\!\!\!\!\!\!\!\!\!\!\!\!\!\!\!\!\!\!\!\!\!\!\!\!\!\!\!\!\!\!\!\!\!\!\!\!\!\!\!\!\!\!\!\!\!\!\!\!\!\!\!\!
& a_1&\frac{(\vf')^4}{\vf^4}+ 2a_2[\frac{(\vf'')^2}{\vf^2} +
\frac{\vf'\vf'''}{\vf^2} -5\frac{(\vf')^2\vf''}{\vf^3}
+\nonumber\\ &+& 3\frac{(\vf')^4}{\vf^4} -
\frac{a''}{a}\frac{(\vf')^2}{\vf^2}+\frac{(a')^2}{a^2}\frac{(\vf')^2}{\vf^2}-
2\frac{a'}{a}\frac{\vf'\vf''}{\vf^2}+ \nonumber\\&+&
2\frac{a'}{a}\frac{(\vf')^3}{\vf^3}]
 -4b[\frac{a''''}{a}-4\frac{a'a'''}{a^2}+6\frac{(a')^2a''}{a^3}+  \nonumber\\&-&3\frac{(a'')^2}{a^2}]
 +24b'\lb\frac{(a')^2a''}{a^3}-\frac{(a')^4}{a^4}\rb  -\frac{24}{\kappa}a^4\La-   \nonumber\\ &-&\frac{12}{\kappa}aa''
 +3\beta a^6(\vf')^2+4a^4V(\vf)=0  \nonumber\\   { }
& &
  \nonumber\\
& 6&a_1\ln a\lb\frac{(\vf')^4}{\vf^5}-\frac{(\vf')^2\vf''}{\vf^4}\rb - 2a_1\frac{a'}{a}\frac{(\vf')^3}{\vf^4}+  \nonumber\\
& +& a^4\frac{\pdr V(\vf) }{\pdr\vf}-\frac{\beta}{2}a^6\vf''-3\beta a^6\frac{a'}{a}\vf'+a_2[\frac{a'a''}{a^2}\frac{\vf'}{\vf^2}+ \nonumber\\
& +&\frac{a''}{a}\lp \frac{(\vf')^2}{\vf^3}
-\frac{\vf''}{\vf^2}\rp - \frac{\vf'}{\vf^2}\frac{a'''}{a}]=0
\eeqs
Here prime means derivative with respect to $y$. In order to
analyze these equations it is interesting to consider the more
simple case, when there is no dilaton. In such a situation from
equations (\ref{ecmov}) it is very easy to obtain the following
equation of motion,
\beqs\la{movc}\!\!\!\!\!\!\!\!\!\!\!\!\!\!\!\!\!\!\!\!
&&\frac{a''''}{a}-4\frac{a'a'''}{a^2}-3\frac{(a'')^2}{a^2}+
\Big(6-6\frac{b'}{b}\Big)\frac{a''(a')^2}{a^3}+\nonumber\\
\!\!\!\!\!\!\!\!\!\!\!\!\!\!\!\!\!\!\!\!&&+\frac{6b'(a')^4}{ba^4}-\frac{a}{4bk}\Big(-12a''-24\La
a^3\Big)=0. \eeqs
So as a result we have obtained the same equation, which is
discussed in \cite{brevik} in the similar situation when the terms
depending on the dilaton are not present. Doing the same analysis,
as in that reference, we see that looking for the special AdS-like
solutions of (\ref{movc}) in the form: $a=c/y$, when there are no
quantum corrections, there is a solution with
$c=1/\sqrt{-\Lambda}$, in accordance with the AdS metric. On the
other hand if the effective cosmological constant $\Lambda=0$,
Eq.~(\ref{movc}) reduces to $c^2=b'\kappa$ (at $a(y)=c/y$).
However, $c^2=b'\kappa$ leads to an imaginary scale factor $a$,
because $b'<0$. We must understand by this that an Anti-de Sitter
Universe can not be created just as a result of matter quantum
corrections. But we know that there exists the possibility of
creation of a de Sitter Universe by solely matter quantum effects
\cite{Starobinskyetc}. Truly, for a scale factor, which depends
only on time, the sign of curvature (which is positive) is
changing. As a result, $a=c/\eta$ with $c^2=-b'\kappa$. So we
conclude that, there is always a solution which leads to a quantum
created de Sitter Universe. On the contrary, we see that a
necessary condition for the existence of an Anti-de Sitter
Universe  is the presence, in classical theory, of a negative
effective cosmological constant.

In the general case for $c^2$ (assuming $\Lambda<0$) it is easily
obtained the following algebraic equation :
\be\kappa b'-c^2-\Lambda c^4=0\;, \ee
and its solutions have the form:
 \be {c_1}^2=-{1\over
2\Lambda}\left(1+\sqrt{1+4\kappa b'\Lambda}\right) \ee
and \be {c_2}^2=-{1\over 2\Lambda}\left(1-\sqrt{1+4\kappa
b'\Lambda}\right). \ee
From these solutions we see that in the first of them, if we start
from some bare (very small) negative cosmological constant, we get
an Anti-de Sitter Universe with a smaller cosmological constant
due to quantum corrections. So we find that the quantum
corrections act against the existing Anti-de Sitter Universe and
make it less stable. This is the mechanism of annihilation of
Anti-de Sitter Universe \cite{brevik}.

The second solution from the physical point of view it is not
interesting because it corresponds to the imaginary scale factor
since in it $c^2<0$.

With the solution like $a(y)=c/y$ it is possible to obtain the
the dependence of the  complete effective action on $c^2$, and we
get for it,
 \beqs &\!\!\!&\!\!\!S_{\rm cl}+W
=\nonumber\\&\!\!\!&\!\!\!= V_3\int{\rmd y\over y^4}{\left[
6b'\ln\!\left({c^2\over y^2}\right)-8(b+b')- {6(\Lambda
c^4-c^2)\over\kappa}\right]}
\nonumber\\
&\!\!\!&\!\!\!= V_3\int{\rmd y\over y^4}{\left[
6b'\ln\!\left({c^2\over y^2}\right)-8b-14b'
+{12c^2\over\kappa}\right]}\;. \eeqs
Summing, because of quantum corrections an already existing
Anti-de Sitter Universe becomes less stable, whereas creation of
an Anti-de Sitter Universe by solely quantum corrections as we
have seen is impossible. Now, one may wonder which effects can be
obtained if we include in our scenario the terms with the dilaton.

\section{Dilaton coupled theories and stabilization of Anti-de Sitter Universe}
As examples of theories where dilaton appears we may consider the
equations (\ref{ecmov}) and look for their solutions for different
dilatonic potentials. It is very interesting to think about the
most common functions for the dilatonic potential (like in the
superstring theories) and look for the possibility of the creation
of an Anti-de Sitter Universe in such a gravitational background.
In the following, for simplicity, we will consider the case when
the kinetic term for the dilaton is absent.

1. Firstly let us see the case when in (\ref{scl}) the dilatonic
potential is not present. In this situation the equations of
motion (\ref{ecmov}) take the form,
\beqs
 && a_1\frac{(\vf')^4}{\vf^4}+ 2a_2[\frac{(\vf'')^2}{\vf^2} +
\frac{\vf'\vf'''}{\vf^2}
-5\frac{(\vf')^2\vf''}{\vf^3} +\nonumber\\
&&+ 3\frac{(\vf')^4}{\vf^4} -
\frac{a''}{a}\frac{(\vf')^2}{\vf^2}+\frac{(a')^2}{a^2}\frac{(\vf')^2}{\vf^2}-
2\frac{a'}{a}\frac{\vf'\vf''}{\vf^2}+ \nonumber\\
&&+ 2\frac{a'}{a}\frac{(\vf')^3}{\vf^3}]
 -4b[\frac{a''''}{a}-4\frac{a'a'''}{a^2}+6\frac{(a')^2a''}{a^3}+  \nonumber \\ &&-3\frac{(a'')^2}{a^2}]
 +24b'\lb\frac{(a')^2a''}{a^3}-\frac{(a')^4}{a^4}\rb  -\frac{24}{\kappa}a^4\La-   \nonumber\\ &&-\frac{12}{\kappa}aa''
 =0  \nonumber\\
&&
  \nonumber\\
&&6a_1\ln a\lb\frac{(\vf')^4}{\vf^5}-\frac{(\vf')^2\vf''}{\vf^4}\rb - 2a_1\frac{a'}{a}\frac{(\vf')^3}{\vf^4}+  \nonumber\\
&&\!\!\!\!\!\!\!\!\!\!\!\!\!\!\!+
a_2[\frac{a'a''}{a^2}\frac{\vf'}{\vf^2}+ \frac{a''}{a}\lp
\frac{(\vf')^2}{\vf^3} -\frac{\vf''}{\vf^2}\rp -
\frac{\vf'}{\vf^2}\frac{a'''}{a}]=0\la{ecmov1} \eeqs
Solving these equations, the first we will do is to make the
transformation to cosmological time (for a review of the method
see \cite{brevik99}): $dz=a(y)dy$. After this, the equations
(\ref{ecmov1}) take the form:
\beqs
 && -4b[a^3\stackrel{....}{a}+3a^2\dot{a}\stackrel{...}{a} + a^2(\ddot{a})^2-5a(\dot{a})^2\ddot{a}]+\nonumber\\
 && +24b'a\dot{a}^2\ddot{a}-\frac{24}{\kappa}a^4\La - \frac{12}{\kappa}\lp a^3\ddot{a}+a^2\dot{a}^2 \rp +\nonumber\\
&&+a_1 a^4\frac{\dot{\vf}^4}{\vf^4}+ 2a_2a^2[a^2\frac{\ddot{\vf}^2}{\vf^2}+\nonumber\\
&&+3a\dot{a}\frac{\dot{\vf}\ddot{\vf}}{\vf^2}+a^2\frac{\dot{\vf}\stackrel{...}{\vf}}{\vf^2} - 5a^2\frac{\dot{\vf}^2\ddot{\vf}}{\vf^3}-3a\dot{a}\frac{\dot{\vf}^3}{\vf^3}+\nonumber\\
&&+3a^2\frac{\dot{\vf}^4}{\vf^4}]=0\nonumber\\
&&\nonumber\\
&& 6a_1\ln a a^2\Big(
a\frac{\dot{\vf}^4}{\vf^5}-a\frac{\dot{\vf}^2\ddot{\vf}}{\vf^4}-\frac{\dot{\vf}^3}{\vf^4}\Big)
-2a_1a^2\dot{a}\frac{\dot{\vf}^3}{\vf^4}+\nonumber\\
&&+a_2( a^2\ddot{a}\frac{\dot{\vf}^2}{\vf^3}-a^2\ddot{a}\frac{\ddot{\vf}}{\vf^2}+ a\dot{a}^2\frac{\dot{\vf}^2}{\vf^3}-a\dot{a}^2\frac{\ddot{\vf}}{\vf^2}+\nonumber\\
&&-a^2\stackrel{...}{a}\frac{\dot{\vf}}{\vf^2}-4a\dot{a}\ddot{a}\frac{\dot{\vf}}{\vf^2}-
\dot{a}^3\frac{\dot{\vf}}{\vf^2}) =0 \la{eqmovz}\eeqs
 Here
$\dot{a}=\rmd a/\rmd z$ and $\dot{\vf}=\rmd \vf/\rmd z$.

As we see these equations are too complicated to be solved
analytically and that is why it is necessary to look for their
solutions approximately. So let us look for following Ansatz:
\beqs a(z)\simeq a_{\scriptscriptstyle0}\,\rme^{Hz}\;,\qquad
\varphi(z)\simeq\varphi_{\scriptscriptstyle0}\,\rme^{-\alpha
Hz}\;.\la{aprox}\eeqs
Taking into account that the logarithmic term  is too small as
$\ln a\sim Hz$ and $H$ is proportional to the Plank mass, we may
neglect it. So after this we obtain from the second equation in
(\ref{eqmovz}) for $\ag$ the following algebraic equation, \beqs
2a_1\ag^3+6a_2\ag =0\;,\eeqs and the solutions for this equation
have the form: \beqs \ag_1 =0\;,\quad \ag_{2,3} = \pm
\sqrt{-\frac{3a_2}{a_1}}=\pm 2i\;.\eeqs
The imaginary solutions are not interesting because, as it is well
known, in that case the creation of an AdS Universe is not
possible as it would be unstable. The trivial solution leads to
the case when there is no dilaton. We may conclude that the
quantum correction we have included in the action, with out
dilatonic potential does not give any effects in the early AdS
Universe.

2. Now let us choose the dilatonic potential in (\ref{scl}) to be
of the form $V(\vf)=\ag\ln \vf$. Performing similar analysis as in
the case before, we obtain the equations of motion in terms of
$z$ in the following form.
\beqs
 && -4b[a^3\stackrel{....}{a}+3a^2\dot{a}\stackrel{...}{a} + a^2(\ddot{a})^2-5a(\dot{a})^2\ddot{a}]+\nonumber\\
 && +24b'a\dot{a}^2\ddot{a}-\frac{24}{\kappa}a^4\La - \frac{12}{\kappa}\lp a^3\ddot{a}+a^2\dot{a}^2 \rp +\nonumber\\
&&+a_1 a^4\frac{\dot{\vf}^4}{\vf^4}+ 2a_2a^2[a^2\frac{\ddot{\vf}^2}{\vf^2}+\nonumber\\
&&+3a\dot{a}\frac{\dot{\vf}\ddot{\vf}}{\vf^2}+a^2\frac{\dot{\vf}\stackrel{...}{\vf}}{\vf^2} - 5a^2\frac{\dot{\vf}^2\ddot{\vf}}{\vf^3}-3a\dot{a}\frac{\dot{\vf}^3}{\vf^3}+\nonumber\\
&&+3a^2\frac{\dot{\vf}^4}{\vf^4}]+4\ag a^4\ln\vf=0\nonumber\\
&&\nonumber\\
&& 6a_1\ln a a^2\Big(
a\frac{\dot{\vf}^4}{\vf^5}-a\frac{\dot{\vf}^2\ddot{\vf}}{\vf^4}-\frac{\dot{\vf}^3}{\vf^4}\Big)
-2a_1a^2\dot{a}\frac{\dot{\vf}^3}{\vf^4}+\nonumber\\
&&+a_2( a^2\ddot{a}\frac{\dot{\vf}^2}{\vf^3}-a^2\ddot{a}\frac{\ddot{\vf}}{\vf^2}+ a\dot{a}^2\frac{\dot{\vf}^2}{\vf^3}-a\dot{a}^2\frac{\ddot{\vf}}{\vf^2}+\nonumber\\
&&-a^2\stackrel{...}{a}\frac{\dot{\vf}}{\vf^2}-4a\dot{a}\ddot{a}\frac{\dot{\vf}}{\vf^2}-
\dot{a}^3\frac{\dot{\vf}}{\vf^2}) +\frac{\ag}{2}\frac{a^3}{\vf}=0
\la{eqmovz2}\eeqs
As it was done above we look for approximated solutions like
 \beqs
a(z)\simeq a_{\scriptscriptstyle0}\,\rme^{Hz}\;,\qquad
\varphi(z)\simeq\varphi_{\scriptscriptstyle0}\,\rme^{-\alpha
Hz}\;.\eeqs
Substituting in the second of equations
(\ref{eqmovz2}) and taking in it  $\ag_{\scriptscriptstyle 0}$
small enough, we obtain for $\ag_{\scriptscriptstyle 0}$ the
following algebraic equation,
\be 4\ag_{\scriptscriptstyle
0}+8(4\pi)^2\ag=0\;,\ee and \be \ag_{\scriptscriptstyle
0}=-32\pi^2\ag\;.\ee
Now from the first equation of
(\ref{eqmovz2}), neglecting the logarithmic term as above, for
$H^2$ the following solution is obtained
\be H^2=-\frac{1}{2\kappa b'}\lp -1+ \sqrt{1+4\kappa
b'\La}\rp\la{H2}\;.\ee
This solution is always positive, and it is important to remark
that in such approximation the solution for $H^2$ does not depend
on the value of parameter $\ag$, and it means that the quantum
correction does not enter into $H^2$. Thus, we have found a non
imaginary scale factor for dilatonic AdS Universe and of course
there occurs the possibility for quantum creation of dilatonic AdS
Universe.

3. Let us see  one more case with the dilatonic potential defined
as $V(\vf)=\ag_1\vf$. Substituting this form for the potential in
(\ref{ecmov}) and performing similar analysis we obtain the
following solution for $\ag_{\scriptscriptstyle 0}$:
\be \ag_{\scriptscriptstyle 0}= \frac{a_{\scriptscriptstyle
0}\ag_{\scriptscriptstyle 1}}{6\kappa a_{\scriptscriptstyle
2}}\rme^{Hz} \;,\ee
and finally for $H^2$ the solution is:
\be H^2=-\frac{1}{2\kappa b'}\big[-1+\sqrt{1+4\kappa \La
b'+4\kappa b'\ag_{\scriptscriptstyle 1}\vf_{\scriptscriptstyle
0}}\big]\;.\ee
This solution is positive, as the third term in the radical is
quite small, so the obtained value for $H^2$ fills the condition
to obtain a non imaginary scale factor and therefore the
possibility for the quantum creation of dilatonic Ads Universe.

\section{Summary}

Summarizing, it was shown that without dilaton the quantum effects
lead to the annihilation of AdS Universe. We have seen how the
form of the dilatonic potential affects the quantum creation of
AdS Universe. It was shown the role of dilaton coupled quantum
matter to stabilization of AdS Universe.

Using anomaly induced effective action for dilaton coupled scalar
and adding it to the classical gravity action, it was done the
analysis of the equations of motion, for a number of dilatonic
potentials. On the basis of this it was shown that the possibility
of quantum creation of dilatonic AdS space occurs.

The interest to AdS space is caused mainly by AdS/CFT
correspondence and nice supersymmetric properties of this
background. There was recently much interest also to brane-world
approach where observable universe is considered as brane in
higher dimensional space. There exits some variant of brane-world
which is induced by brane quantum effects (so-called  Brane New
World)(for its formulation see \cite{NOZ,Hawking}).

It would be very interesting to understand the role of choice of
dilatonic potential to quantum creation of dilatonic de Sitter and
Anti-de Sitter spaces in Brane New World scenario. This will be
discussed elsewhere.

 \Acknow {We are  grateful to Prof.S.D. Odintsov for
formulation the problem and numerous helpful discussions. We thank
Prof.P.M. Lavrov for very useful discussions. The research by
J.Q.H.  was supported in part by Professorship and Fellowship from
the Universidad Tecnologica de Pereira, Colombia.}

\end{document}